\documentclass[12pt]{article}
\pdfoutput=1
\usepackage{amsfonts,amsthm,amscd,mathrsfs,helvet}
\usepackage{amsbsy}
\usepackage[format=hang]{caption}
\usepackage{epsfig,amssymb,amsmath,graphicx,subcaption,verbatim,hyperref,xcolor,
ulem,epstopdf,psfrag,pstool,braket,array,enumerate,multirow,wrapfig}
\usepackage{graphics,color}
\newcommand{\be}{\begin{equation}}
\newcommand{\ee}{\end{equation}}
\newcommand{\bea}{\begin{eqnarray}}
\newcommand{\eea}{\end{eqnarray}}
\newcommand{\bean}{\begin{eqnarray*}}
\newcommand{\eean}{\end{eqnarray*}}
\font\upright=cmu10 scaled\magstep1
\font\sans=cmss12
\newcommand{\ssf}{\sans}
\newcommand{\stroke}{\vrule height8pt width0.4pt depth-0.1pt}
\newcommand{\Z}{\hbox{\upright\rlap{\ssf Z}\kern 2.7pt {\ssf Z}}}

\newcommand{\C}{{\rlap{\rlap{C}\kern 3.8pt\stroke}\phantom{C}}}
\newcommand{\R}{\hbox{\upright\rlap{I}\kern 1.7pt R}}
\newcommand{\CP}{\C{\upright\rlap{I}\kern 1.5pt P}}
\newcommand{\identity}{{\upright\rlap{1}\kern 2.0pt 1}}
\newcommand{\half}{\frac{1}{2}}

\begin{document}
\pagestyle{plain}

\title{\vskip -70pt
\begin{flushright}
{\normalsize DAMTP-2017-29} \\
\end{flushright}
\vskip 50pt
{\bf \LARGE \bf Lightly Bound Skyrmions, Tetrahedra and Magic Numbers}
 \vskip 30pt
}
\author{{\bf Nicholas S. Manton}
\thanks{email: N.S.Manton@damtp.cam.ac.uk} \\[15pt]
{\normalsize
{\sl Department of Applied Mathematics and Theoretical Physics,}}\\
{\normalsize {\sl University of Cambridge,}}\\
{\normalsize {\sl Wilberforce Road, Cambridge CB3 0WA, England.}}\\
}
\vskip 20pt
\date{July 2017}
\maketitle
\vskip 20pt

\begin{abstract}

In the lightly bound Skyrme model, several Skyrmions having particularly 
strong binding are clusters of unit baryon number Skyrmions 
arranged as truncated 
tetrahedra. Their baryon numbers form the sequence $B = 4 \,, 16 \,, 40 \,, 
80 \,, 140 \,, 224$. This is the standard sequence of tetrahedral numbers
multiplied by four, and therefore agrees with the sequence of magic proton 
and neutron numbers $2 \,, 8 \,, 20 \,, 40 \,, 70 \,, 112$ that occurs 
in the nuclear shell model in the absence of strong spin-orbit coupling. This 
sequence includes several of the magic numbers that are predicted 
for tetrahedrally deformed nuclei, and also appears in the context
of the FCC lattice geometry investigated long ago by Wigner and revived
more recently. 

\end{abstract}

\vskip 30pt

\newpage

\section{Introduction}

The Skyrme model proposes that baryons are topological solitons in a
field theory of pions \cite{Sk,RZ}. Protons and neutrons are spin-half quantum
states of the basic Skyrmion with unit baryon number, and they combine into 
an isospin-half doublet of nucleons \cite{ANW}. The model incorporates 
approximate chiral symmetry and has
had considerable success in modelling not only nucleons, but also the 
ground states and excited states of larger nuclei. Many Skyrmion
solutions with higher baryon numbers are known. The simplest
quantization technique is rigid body quantization of the classical
Skyrmions of minimal energy, but recent work has considered 
some of the low-energy deformation modes of Skyrmions; this gives further
states, and improved fits to the spectra of nuclei, in particular
Carbon-12 \cite{LM} and Oxygen-16 \cite{HKM}.

However, two problems with the standard Skyrme model are that the classical 
Skyrmion binding energies are rather high, and that there is little evidence
for a conventional sequence of magic numbers that should be a feature
of a model of nuclei.

There have been various attempts to resolve the first problem by
devising variants of the standard Skyrme model, and the variant we 
consider here is the one devised by the group in Leeds \cite{Har,GHS,GHK}. 
This adds a quartic potential term to the usual pion mass term,
and also substantially changes the ratio between the quadratic and
quartic terms in field derivatives (the sigma model term and
Skyrme term). The model has much reduced classical binding energies,
although this desirable feature is rather spoiled by quantum effects. 

The Leeds group found the classical energy minima for baryon
numbers\footnote{We use the notation $B$ for baryon number, as is
standard in Skyrmion research. It is identical to atomic
mass number, usually denoted by $A$.}
up to $B = 8$ in \cite{GHS}. In addition to the low binding energies, their
most striking result was that the Skyrmions were clusters of almost
undeformed $B=1$ Skyrmions centred very close to vertices of a face centred 
cubic (FCC) lattice. In \cite{GHK} they showed that an accurate approximation 
to their model is obtained by treating the $B=1$ Skyrmions
as particles located precisely on the FCC vertices. They
could study Skyrmions up to baryon number $B=23$ using this approximation. 

It was not really a novel discovery that the optimal way to arrange
large numbers of $B=1$ Skyrmions in three dimensions 
is to place them at the vertices of an FCC lattice. This was first
noted by Kugler and Shtrikman \cite{KSh}, and Castillejo et al. \cite{CJJ}. 
A similar lattice of instantons probably occurs in the holographic
approach to multi-baryon systems \cite{MS,KMS}.
The FCC lattice has four sublattices, and the Skyrmions have a uniform 
orientation on each sublattice. Nearest neighbour Skyrmions, which are 
on distinct sublattices, have a relative orientation that is maximally 
attractive (i.e. one Skyrmion is rotated relative to the other by $\pi$ around
a line perpendicular to the line joining them), and this is why the 
overall energy is minimised. 

In the standard Skyrme model with zero pion mass one can find an infinite, 
periodic Skyrme crystal. If the lattice spacing is forced to be relatively 
large, the crystal structure is an FCC lattice of $B=1$ Skyrmions,
but as the lattice spacing decreases, it reaches a critical value
where the Skyrmions partially merge and the symmetry is enhanced. 
The crystal then becomes a
primitive cubic lattice of half-Skyrmions. The true energy minimum of
the Skyrme model occurs at a lattice spacing smaller than the
critical value, so it is a crystal of half-Skyrmions. At least, this
is the case if one just considers the static solutions for zero pion mass. The 
half-Skyrmion crystal structure is remarkable, but for massive pions 
the enhanced symmetry (a kind of discrete chiral symmetry) is lost and
the minimal energy crystal reverts to having FCC structure. The FCC 
lattice is also almost certainly the minimal energy crystal 
structure in the lightly bound Skyrme model. 

Ma and Rho \cite{MR} have recently reconsidered the transition between 
the FCC crystal of Skyrmions and the half-Skyrmion crystal with enhanced
symmetry. By taking into account pion mass effects and quantum
effects, they argue that at normal nuclear densities, the Skyrmions
are in the FCC phase, but at higher densities (more than twice normal
nuclear densities) the half-Skyrmion phase will occur. This has
consequences for neutron stars and other dense nuclear systems that
are not normally accessible in laboratory experiments. 
It suggests that it is reasonable to study variants of the
Skyrme model with FCC arrangements of $B=1$ Skyrmions, where the
Skyrmions are close to merging. The lightly bound model is just one
such variant.

Related to the transition from the FCC crystal to the half-Skyrmion
crystal is what happens for baryon number
$B=4$. The optimal way to arrange four $B=1$ Skyrmions is to place them
at the vertices of a regular tetrahedron. The orientations are
distinct and are those that occur on the four FCC sublattices. All six pairs
of Skyrmions maximally attract, and the field configuration has tetrahedral
symmetry. Remarkably, in the standard Skyrme model, both for zero mass 
pions and for pions of realistic mass, the true $B=4$ Skyrmion of minimal 
energy has an enhanced cubic symmetry \cite{BTC}. The four Skyrmions at the 
vertices of the tetrahedron merge into a cubic structure with 
half-Skyrmions at the eight vertices. However, for the lightly bound
model, and probably many other variant models, the minimal energy
solution remains tetrahedral. Note that a cubic structure easily deforms 
into a tetrahedral structure, and in two ways, because the vertices of a cube
naturally split into two subsets forming tetrahedra. The relevant 
tetrahedral symmetry group is ${\rm T}_d$, a subgroup of 
the cubic group ${\rm O}_h$. It is therefore not surprising that the 
$B=4$ Skyrmion has a vibrational mode that oscillates between two
dual tetrahedra, and that this is one of the lowest frequency modes \cite{BBT}.

The existence of the cubic $B=4$ Skyrmion in the standard Skyrme model
has influenced much of the recent research into Skyrmions of higher
baryon numbers. Skyrmion solutions, for baryon numbers a multiple of 
four, have been found by bringing several $B=4$ cubes together \cite{BMS}. The
results are similar to those found in the alpha-cluster models of
nuclei. The Skyrmion with $B=8$ has two cubes touching along a face.
Solutions with $B=12$ have been found with three cubes arranged in an
equilateral triangle, and also in a linear chain; these have similar
energies. Four $B=4$ cubes can be arranged tetrahedrally to give 
a $B=16$ solution. Eight cubes can be arranged into a large cube with $B=32$,
and twenty seven $B=4$ cubes produce the largest known standard Skyrmion, with
$B=108$ \cite{FLM}. 

However, the study of higher baryon number Skyrmions as clusters of
$B=4$ cubes, in the standard Skyrme model, has reached an impasse. 
Quantizing the cubic $B=32$ and $B=108$ 
Skyrmions as rigid bodies doesn't work well. It has also not been possible to
construct a $B=40$ solution from ten $B=4$ cubes, to obtain a good
model for the magic nucleus Calcium-40. More generally, the standard
Skyrme model has not yet yielded obvious structures compatible with the
known magic numbers for nuclei, beyond the $B=16$ solution modelling 
Oxygen-16. 

More promising, then, is the lightly bound Skyrme model, with its symmetries
inherited from the FCC lattice. The maximal symmetry of clusters cut
out from the FCC lattice is cubic, but these cubically symmetric
clusters are not exceptionally tightly bound. The most tightly bound
clusters appear to have tetrahedral symmetry, and have a single tetrahedral
cluster of four $B=1$ Skyrmions at their centre. We shall describe
these next. Their baryon numbers match the magic numbers established 
in other nuclear models.

\section{Clusters with Tetrahedral Symmetry}

In the FCC lattice, each vertex has 12 nearest neighbours. Equivalently, the
coordination number is 12. In the lightly bound Skyrme model there is
one baryon (i.e. one $B=1$ Skyrmion) at each vertex, and it is an 
accurate approximation to say that the binding is predominantly due 
to the pair interactions between nearest neighbours, which are all of 
the same strength. We shall use this approximation, and 
ignore longer-range contributions to the interactions. We refer to 
the binding between each nearest neighbour
pair as a bond, so the total binding energy is a constant times
the number of bonds. Let us normalise the energy so that this constant is
unity, and identify the number of bonds as the binding energy $E$.

Using this approach, we see that as each baryon in the FCC lattice
is bonded to twelve others, and each bond has two baryons at its ends, 
the binding energy per baryon of the complete lattice is $E/B = 6$. We are
interested in finite clusters of baryons arranged as subsets of the 
FCC lattice. $E/B$ is then obviously less than 6, 
because some baryons, especially those on the surface of a cluster, 
have fewer than twelve nearest neighbours. 

The smallest clusters beyond a single pair have baryon numbers $B=3$ and
$B=4$. The $B=3$ cluster is an equilateral triangle of baryons, with
3 bonds and $E/B = 1$; the $B=4$ cluster is a tetrahedron of baryons, with
6 bonds and $E/B = 1.5$. The next highly symmetric cluster is the
$B=6$ octahedron with 12 bonds, for which $E/B = 2$. For baryon numbers
less than 16 it is not possible for $E/B$ to be as large as 3. There are
highly symmetric clusters with $B=13$ (a single baryon surrounded by
all twelve of its nearest neighbours) and $B=14$ (a $B=6$ octahedron with
each face completed into a tetrahedron by adding one more
baryon outside). These are both cubically symmetric, and have 36 and 40 bonds
respectively, giving $E/B$ values of 2.77 and 2.86. Notice that the
cubic symmetry is differently realised in these two cases, as the
first cluster has a baryon at its centre, but the second does not.

For $B=16$ there is a tetrahedrally symmetric cluster of baryons with 48 bonds,
so $E/B = 3$. This has the basic $B=4$ tetrahedron at the
centre, attached to a triangular $B=3$ cluster above each face. It 
therefore consists of four $B=6$ octahedra each sharing one
face with the central tetrahedron, and these octahedra have some shared
edges and vertices. Four outer faces are hexagons with seven
baryons, and four are triangles with three baryons (see Figure 1). 

\begin{figure}[!ht]
\centering
\includegraphics[width=10.5cm]{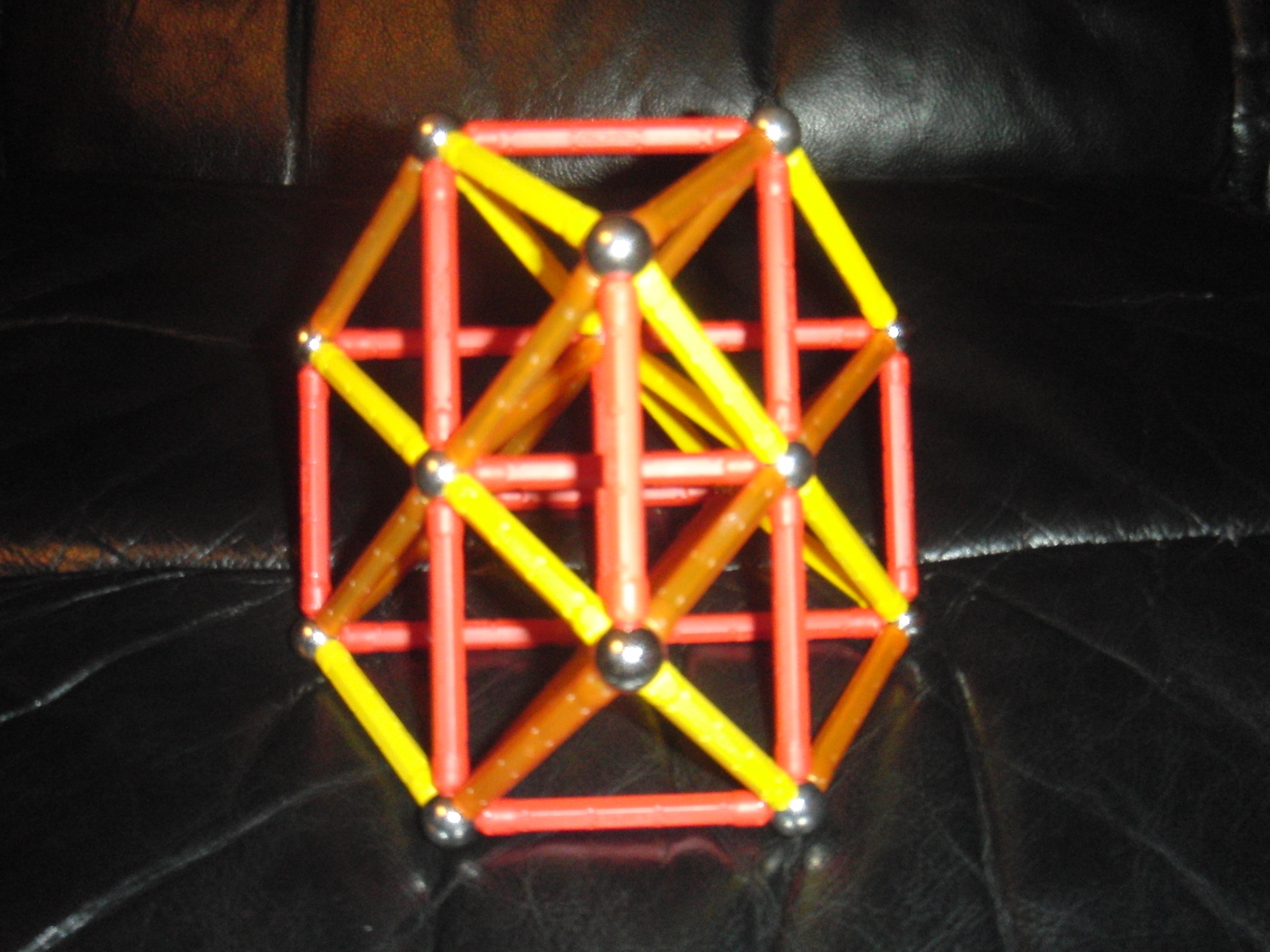}
\caption{$B=16$ cluster in the lightly bound Skyrme model.}
\label{fig1}
\end{figure}

Notice that this cluster inherits a basic property of the FCC
lattice. The complete lattice can be decomposed into alternating 
$B=4$ tetrahedra and $B=6$ octahedra. Each tetrahedron shares
faces with four neighbouring octahedra, and each octahedron shares
faces with eight neighbouring tetrahedra. The octahedra all have 
the same spatial orientation, whereas the tetrahedra occur in two 
orientations related by inversion. The $B=16$ cluster has a single
central tetrahedron, together with its four neighbouring octahedra and 
six more tetrahedra between the octahedra. (The $B=14$
cluster mentioned above has a single central octahedron together with its
eight neighbouring tetrahedra.)

The larger clusters we shall consider are mainly those built on the $B=16$
core, and the most symmetric ones have tetrahedral symmetry. One
can also consider larger clusters with cubic symmetry, built on the
$B=13$ or $B=14$ cores. For a discussion of these, and some further
clusters, see the Appendix. The tetrahedral clusters have large $E/B$ 
values, but they have competitors. For example, at $B=19$, there is a large
octahedron (with a baryon at the centre) with 60 bonds; there is also
a cluster where a $B=3$ triangle is attached to one of the hexagonal
faces of the $B=16$ cluster, also with 60 bonds, but less symmetry. 
It is the second cluster that easily allows further baryons to be attached, 
so as to achieve higher $E/B$ values.   

At this point it is helpful to introduce Cartesian coordinates for the vertices
of the FCC lattice. We choose the origin to be the centre of one of
the $B=4$ tetrahedra, and orient and scale this tetrahedron so that its
vertices are at $(1,1,1)$, $(1,-1,-1)$, $(-1,1,-1)$ and $(-1,-1,1)$. 
The vertices of the entire FCC lattice are then at the positions 
$(a,b,c)$ where $a$, $b$ and $c$ are odd integers, and also $a+b+c = 3
\bmod 4$. The vectors from any vertex to its nearest neighbours 
are $(0,\pm 2,\pm 2)$, $(\pm 2,0,\pm 2)$ and $(\pm 2,\pm 2,0)$, and 
have squared length 8. Note that the sum of the coordinates of these 
vectors is always $0 \bmod 4$. The points $(a,b,c) = (1,1,1) \bmod 4$ 
form one of the four sublattices of the FCC lattice, and $B=1$ Skyrmions
located at these points all have the same orientation. Similarly for 
the points equal to $(1,-1,-1)$, $(-1,1,-1)$ or $(-1,-1,1) \bmod 4$.    

It is straightforward to classify larger, tetrahedrally symmetric
clusters using these coordinates. For example, the $B=16$ cluster has
baryons at all the allowed vertices with coordinates $(\pm 1,\pm 1,\pm 1)$,
and the vertices whose coordinates are $(\pm 3,\pm 1,\pm 1)$ or
its permutations. The constraint $a+b+c = 3 \bmod 4$ allows
half of the sign combinations here, so there are four vertices of the
first type and twelve of the second type. These are at squared distances 3
and 11 from the origin, respectively. The next largest tetrahedrally symmetric
cluster adds baryons at the twelve allowed vertices with coordinates 
$(\pm 3,\pm 3,\pm 1)$ and its permutations, all at
squared distance 19, to produce $B=28$. This efficiently adds four 
triangular $B=3$ clusters above each hexagonal face of the $B=16$
cluster, and adds 48 bonds, producing 96 bonds in total. (Note 
that the squared distances are always of the form $8k+3$.)    

At squared distance 27 there are two sets of vertices. There are twelve
vertices $(\pm 5,\pm 1,\pm 1)$ and its permutations, and 
four vertices $(\pm 3,\pm 3,\pm 3)$. We discover here that tetrahedral 
symmetry is not a sufficient criterion for achieving a large value of 
$E/B$. It is optimal to add twelve baryons on the first set of vertices,
to produce $B=40$, but not optimal to add four baryons on 
the second set either before or after adding the twelve. The twelve
baryons of the first set occur in six pairs; each pair attaches to two
touching square faces of the $B=28$ cluster, adding 9 bonds per
pair, and 54 bonds altogether. The four baryons of the second set are isolated 
and add only 3 bonds each.

Therefore, there are tetrahedrally symmetric clusters with $B=32$ and 
with $B=44$, but the most strongly bound cluster in this region of 
baryon numbers is the $B=40$ cluster, which has $96 + 54 = 150$ bonds, 
and $E/B = 3.75$ (see Figure 2). It has the form of a truncated tetrahedron, 
as does the $B=16$ cluster. Baryon numbers 4, 16 and 40 are magic numbers,
corresponding to the nuclei Helium-4, Oxygen-16 and Calcium-40, which
is encouraging, so we shall study these truncated tetrahedra further. 
For their baryon numbers, they have maximal or almost maximal
values of $E/B$.  

\begin{figure}[!ht]
\centering
\includegraphics[width=10.5cm]{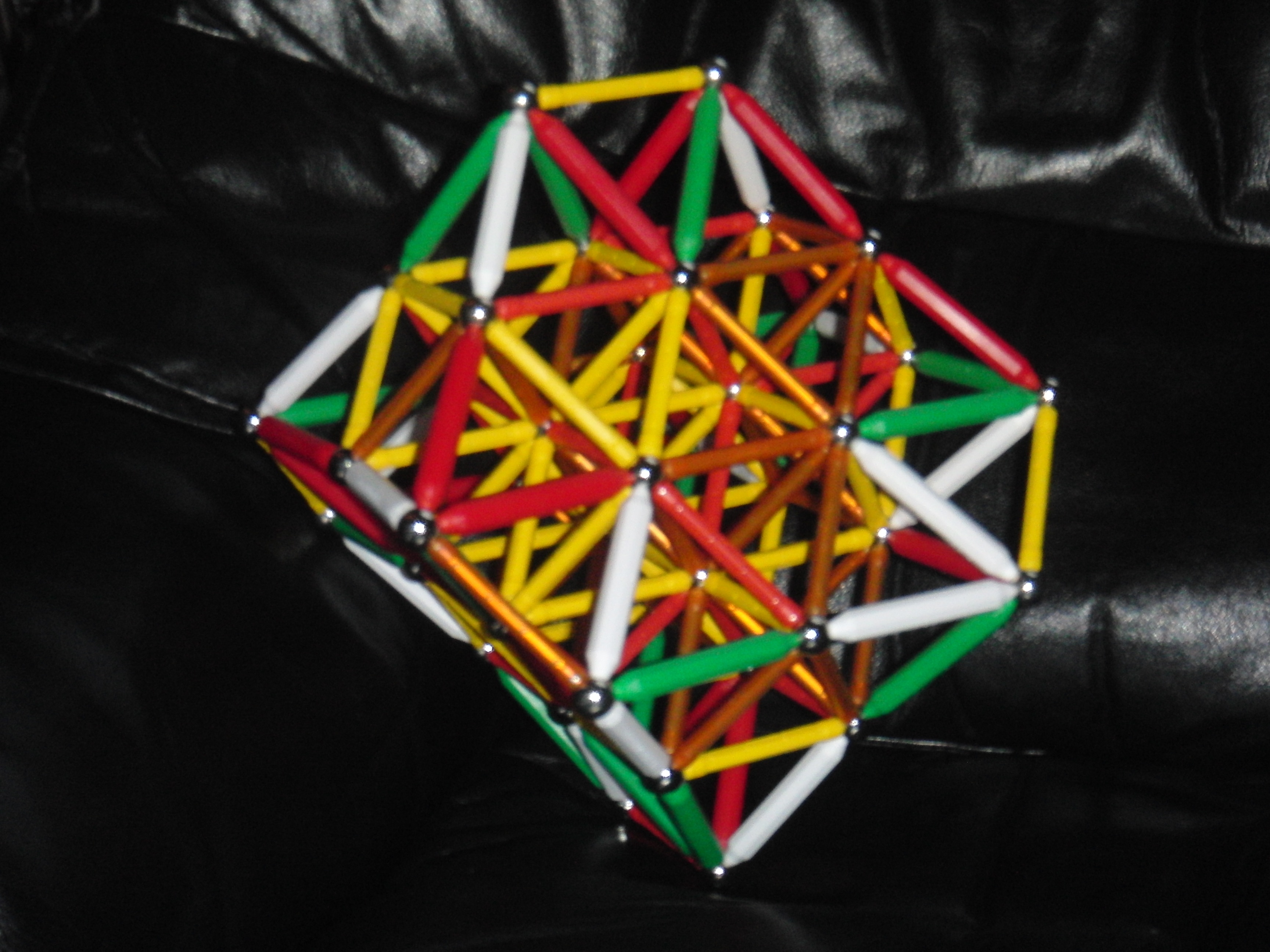}
\caption{Truncated tetrahedral $B=40$ cluster.}
\label{fig2}
\end{figure}

\section{Magic, Truncated Tetrahedra}

We now discuss in more generality the infinite family of truncated 
tetrahedral clusters that give exceptionally large values of the
binding energy per baryon, $E/B$, and shall refer to their baryon 
numbers as magic.

The FCC lattice has complete, pure tetrahedra as subclusters. These are 
obtained by intersecting four planes parallel to the faces of the basic $B=4$ 
cluster at the centre. The first of these clusters beyond $B=4$ has
$B=10$, but this does not have a $B=4$ tetrahedron at its centre,
so more interesting is the next one, with $B=20$. When four baryons
are truncated from its vertices we recover the magic $B=16$ cluster. More 
generally, complete tetrahedra have too much of a pointed shape to have an
exceptional value for $E/B$. What we need to do is to truncate the 
tetrahedra by slicing off four smaller tetrahedra to bring the 
cluster closer to a spherical shape.

To calculate the total baryon numbers of these truncated tetrahedra we
require the pure tetrahedral numbers. A tetrahedron is built up from layers
of equilateral triangles, so a tetrahedral number $T_N$ is the sum of a
sequence of triangular numbers,
\be
T_N = \sum_{n=1}^N \half (n+1)n = \frac{1}{6}(N+2)(N+1)N \,.
\ee
The first few of these are $1 \,, 4 \,, 10 \,, 20 \,, 35 \,,
56 \,, 84 \,, 120 \,, 165 \,, 220 \,, 286 \,,364$.
(An amusing appearance of these numbers is in the song Twelve Days of
Christmas. If one takes literally that on the first day the gift
is a Partridge in a Pear Tree, and on the second it is two Turtle
Doves {\it and} a Partridge in a Pear Tree, and so on, then by the 12th 
day the total number of gifts is $T_{12} = 364$.)

If we require our truncated tetrahedra to have a central $B=4$
cluster, which appears desirable, then 
we must start with a complete tetrahedron whose edge has an even number 
of baryons. We then truncate this by removing four equal tetrahedra of
baryons, leaving the shortest edge with just two baryons. This
produces the truncated structure that is closest to spherical.
We therefore start with a tetrahedron with $2N$ baryons on an edge, 
and remove four tetrahedra with $N-1$ baryons on 
an edge. The total baryon number remaining is
\bea
M_N = T_{2N} - 4T_{N-1} &=& \frac{1}{6}(2N+2)(2N+1)2N -
\frac{2}{3}(N+1)N(N-1) \nonumber \\
&=& \frac{2}{3}(N+1)N(2N+1-N+1) \nonumber \\
&=& \frac{2}{3}(N+2)(N+1)N \,.
\eea
There are a pleasing number of common factors in the two
contributions to $M_N$, and the result,
surprisingly, is four times the tetrahedral number $T_N$. The cluster has
$T_N$ baryons in each of the four orientations, matching the equal
distribution of orientations that occurs for the central $B=4$ 
tetrahedron, but the baryons with a given orientation are not arranged 
as a pure tetrahedron.

The first few of the numbers $M_N$ are $4 \,, 16 \,, 40 \,, 80 \,, 
140 \, ,224$, and we refer to these as tetrahedral magic (baryon) numbers. 
Assuming a nucleus with one of these baryon numbers has equal 
numbers of protons and neutrons, then these form
the sequence $2 \,, 8 \,, 20 \,, 40 \,, 70 \,, 112$. The first three 
of these are the standard magic numbers of nuclear physics. 

We calculate next the total bond numbers of the truncated
tetrahedra. We again start with a complete, pure tetrahedron, and consider
slicing it up into parallel equilateral triangles. A triangle with $n$
baryons along an edge has (the triangular number) $\half n(n-1)$ bonds
in each of three directions, so there are $\frac{3}{2} n(n-1)$ bonds
within this triangle. The triangle is also bonded to the next smaller
triangle by 3 bonds for each baryon in the smaller triangle, so
there are $\frac{3}{2} n(n-1)$ such bonds. This triangle therefore 
contributes $3n(n-1)$ bonds overall, which is six times a triangular number. 
Summing these up, we find that a complete tetrahedron with $N$ baryons
along each edge has $(N+1)N(N-1)$ bonds in total, six times a
tetrahedral number.

Our truncated tetrahedron starts as a
complete tetrahedron with $2N$ baryons on an edge, and then four
tetrahedra with $N-1$ baryons along an edge are removed. The bonds of
the removed tetrahedra are lost, as are the bonds connecting these
tetrahedra to what remains. The total number of bonds of the
truncated tetrahedron is therefore
\bea
E_N &=& (2N+1)2N(2N-1) - 4\left(N(N-1)(N-2) + \frac{3}{2}N(N-1)\right) 
\nonumber \\
&=& 2(2N-1)(N+2)N \,.
\eea
The first few of these bond numbers are $6 \,, 48 \,, 150 \,, 
336 \,, 630 \,, 1056$, and these are also the approximate binding
energies. The binding energies per baryon are $E/B = 1.5 \,, 3 \,, 3.75 \,, 
4.2 \,, 4.5 \,, 4.72$, and the general algebraic formula is
\bea
E/B = E_N/M_N &=& \frac{2(2N-1)(N+2)N}{\frac{2}{3}(N+2)(N+1)N}
\nonumber \\
&=& 3\frac{2N-1}{N+1} \nonumber \\
&=& 6 - \frac{9}{N+1} \,.
\eea
This slowly approaches 6 as expected, but only reaches 5 for $N=8$,
when $B=480$, much larger than the baryon number of any observed nucleus.

An alternative way to find the bond number of a truncated tetrahedron 
is to note that baryons occur in four
types of position. There are interior baryons, face baryons (not 
on an edge), edge baryons (not at a vertex), and vertex baryons. 
These have coordination numbers 12, 9, 7 and 5, respectively. As each
bond has two ends, the total bond number is half of the total 
coordination number. For example, the $B=140$ truncated tetrahedron 
(which has edge lengths of 2 baryons and 5 baryons) has 40 interior 
baryons, 52 face baryons, 36 edge baryons and 12 vertex baryons, and the
total bond number is 630. 

Some of the truncated tetrahedra have another interesting property. 
When $N$ is odd, the truncated tetrahedron has an even number of 
parallel layers with triangular symmetry, and
in this case it can be decomposed into a set of disjoint $B=4$ tetrahedra
that account for all the baryons. These $B=4$ tetrahedra are connected by
octahedra sharing the faces. The interpretation is that the
truncated tetrahedra are clusters of alpha particles. In particular, 
the truncated tetrahedra with baryon numbers 40 and 140 decompose 
into 10 and 35 disjoint $B=4$ tetrahedra, respectively, arranged 
with tetrahedral symmetry. The case $B=40$ is particularly
interesting, because until now it was not known how to arrange ten 
alpha particles into a Calcium-40 nucleus in any Skyrmion model.

The orientations of the $B=4$ tetrahedra alternate. The arrangement
for $B=40$ is that six tetrahedra have one orientation and four the other. The
six occur at vertices of an large octahedron, and the four at vertices of a
large tetrahedron. The ten together do not form a pure tetrahedron. For
$B=140$ there are nineteen in one orientation, at the vertices of a larger
octahedron, and sixteen in the other orientation, at the vertices of a
larger truncated tetrahedron. It is curious that the symmetry group of one
set is cubic, and larger than that of the other set.

\section{Physics of Truncated Tetrahedra}

We propose that truncated tetrahedra composed of lightly bound Skyrmions 
are models for some types of magic nuclei, and present some of
the evidence here. Recall that the magic baryon numbers we have obtained are 
$4 \,, 16 \,, 40 \,, 80 \,, 140 \,, 224$. The first three of these 
clearly correspond to Helium-4, Oxygen-16 and Calcium-40. The 
fourth corresponds to Zirconium-80. This is conjectured to be 
magic, despite being close to the proton drip line, because 40 appears
to be a magic number for both protons and neutrons in tetrahedrally deformed 
nuclei \cite{TYM,DGSM}. Baryon number 140 is only conjecturally magic, based
on the evidence that 70 is a magic number for protons/neutrons in 
tetrahedrally deformed nuclei. However, no nucleus exists with both 
70 protons and 70 neutrons (the largest, short-lived nucleus with equal
proton and neutron numbers is Tin-100, or possibly Tellurium-104). 
Baryon number 224 allows for Radium-224, which is octupole-deformed
and possibly tetrahedral \cite{LD}, with 88 protons and 136 neutrons. 

The binding energy per baryon, eq. (4), matches the two leading terms
in the Bethe--Weizs\"acker, liquid drop mass formula
\be
E/B = a_{\rm V} - a_{\rm S} B^{-\frac{1}{3}} \,,
\ee
where empirically, $a_{\rm S}/a_{\rm V} \simeq 1.1$. Using eq. (2), we
see that it is a very good approximation to write $B \simeq
\frac{2}{3} (N+1)^3$, and then eq. (4) becomes
\be
E/B = 6 - 9\left(\frac{3}{2}B\right)^{-\frac{1}{3}} \,.
\ee
The prediction from the lightly bound Skyrme model is therefore that 
$a_{\rm S}/a_{\rm V} = \left(\frac{3}{2}\right)^{\frac{2}{3}} \simeq 1.3$.

One feature of magic nuclei, according to the shell model, is 
that there is a large energy gap
between the highest filled level and the lowest unfilled level. In
other words, it takes more than the usual energy to excite a single nucleon.
The truncated tetrahedral Skyrmions have an analogous feature. 
For a truncated tetrahedron, the minimal coordination number of a
baryon is 5. This occurs for the baryons at the twelve vertices. For 
other cluster shapes there is usually a baryon with a coordination 
number smaller than this. For example, the baryons at the vertices 
of pure tetrahedra have coordination number 3, and baryons at the 
vertices of untruncated octahedra have coordination number 4. (However, 
there are truncated octahedra where all coordination numbers are 6 
or more -- see Appendix). The energy required to remove 
one baryon from a truncated tetrahedron is therefore large, since 
5 bonds need to be broken, and this is evidence for it being 
magic. Moreover, all the elementary faces of a truncated tetrahedron 
are triangles, so the optimal way to relocate the baryon is to attach 
it by 3 bonds to one of these triangles. Moving 
a single baryon from a vertex to one of these new locations is
therefore at the cost of 2 bonds, so the one-baryon excitation energy 
(the energy of a nuclear particle-hole excitation) is 2 bond units. By
identifying 6 bond units with $a_{\rm V} = 15.6$ MeV, we see that the
bond unit is $2.6$ MeV, so the one-baryon separation energy is
predicted to be 13 MeV, and the one-baryon excitation energy to be
$5.2$ MeV.

It is also interesting to consider what can happen if a few baryons
are added to a truncated tetrahedral core. A single baryon can be 
attached with 3 bonds (as just
mentioned), but two neighbouring baryons can be added with 7 bonds,
and it is rather efficient to attach a triangular cluster of three 
baryons to an underlying face, which adds 12 bonds. This
significantly increases $E/B$ in the case that the core has $B=16$ or $B=40$. 
Attaching a triangular $B=3$ cluster in this way could
provide a model for Fluorine-19 (the only stable isotope of
Fluorine), or Scandium-43 \cite{SW}. The larger faces of the $B=40$ 
core accommodate attaching a
hexagonal cluster of 7 baryons, which adds 33 bonds. This could model 
Scandium-47 or Titanium-47, which are moderately stable compared 
to neighbouring isotopes. Finally, the $B=80$ core accommodates attaching 
a triangular cluster of 10 baryons to a large face, which adds 48
bonds, and could be related to 50 being a magic number.  

The simplest, original version of the shell model of nuclei \cite{BW}
supposed that the individual protons and neutrons move in a mean field
potential that is a three-dimensional isotropic harmonic oscillator. 
The harmonic oscillator energy levels have high degeneracies, and a magic
nucleus is one where all the states up to a given energy are filled. 
Allowing for two spin states, the magic proton and neutron numbers 
are precisely $2 \,, 8 \,, 20 \,, 40 \,, 70 \,, 112$, double the 
tetrahedral numbers. The appearance of tetrahedral numbers here is
well known, but still rather surprising, because the mean field potential is
spherically symmetric. Moreover, there seems no
obvious connection with the spatial structure of a truncated
tetrahedron that we have discussed. 

The refined version of the shell model introduces a more
sophisticated energy-dependence on the orbital angular momentum of
each nucleon, and includes a strong spin-orbit force. The net effect
is to raise the magic numbers $40$, $70$, $112$ to $50$, $82$,
$126$. Magic nuclei do not have to have a magic baryon number. It is
sufficient if either the proton {\it or} neutron numbers are magic.
We do not yet understand this in the context of Skyrmions, but are
still encouraged to find that Skyrmion magic numbers overlap
those of nuclear physics in the cases of equal proton and neutron 
numbers. Particularly encouraging is to find that in the lightly 
bound Skyrme model, the Skyrmion with baryon number 40 is magic, and 
can be used to model Calcium-40. In the standard Skyrme model, a 
$B=40$ solution of the field equations with similar symmetry probably 
exists, and the search for it is underway. 

The shell model usually assumes a spherical mean field potential, but there
have been substantial investigations over many decades of deformed 
shapes. These deformations are usually assumed to be
quadrupolar, producing an ellipsoidal shape with ${\rm D}_{2h}$
symmetry \cite{EG}, but octupolar deformations, including the special 
octahedral deformations that preserve tetrahedral symmetry have also 
been analysed \cite{TYM,DGSM}. The magic numbers that would occur 
for such tetrahedrally deformed shapes appear 
to be much closer to the sequence one finds in the original shell 
model. The tetrahedral deformation
seems to suppress the effect of the spin-orbit force. The magic proton
and neutron numbers for tetrahedral nuclei include 40, 56 and 70. If
one could ignore Coulomb effects, this would lead to magic baryon
numbers 80 and 140, just as for the Skyrmions that are truncated tetrahedra. 
The relation between the (quantum) shell model calculations
exploring tetrahedral deformations and the (classical) spatial 
structures we have found is not clear, despite the commonality of 
an underlying tetrahedral symmetry.
 
In the discussions of tetrahedrally deformed nuclei using the shell
model, there is little consideration of the small proton and neutron magic 
numbers 2, 8 and 20, because the corresponding magic nuclei are usually 
supposed to be intrinsically spherical. However, a small
tetrahedral deformation would not spoil these magic numbers, and
cluster models of nuclei suggest that Oxygen-16 at least has a
tetrahedral form.

Rigid body quantization of a tetrahedral structure leads to a
non-standard rotational band, with spin/parities
$J^P = 0^+,3^-,4^+,6^+,6^-,7^-,...$ (see \cite{Lez1}, for example). 
The existence of appropriately spaced $3^-$ and $4^+$
states, and absence of a lower-lying $2^+$ state is therefore a key
indicator of a tetrahedral structure. Magic nuclei like Oxygen-16,
Calcium-40 and Lead-208 are well known for having low-lying $3^-$ 
states and no such $2^+$ states, and encourage the interpretation 
of these magic nuclei as tetrahedrally deformed, although 
alternative interpretations in terms of octupole vibrations have 
also to be considered. 

Within the standard Skyrme model, there have been a number of
investigations of the quantum states of tetrahedrally and cubically 
symmetric Skyrmions \cite{LM1}. There are constraints on the spin/parities
determined by the symmetry of the Skyrmion and the topological
Finkelstein--Rubinstein sign factors related to the symmetry 
group elements. Provided the baryon number is a multiple of four, and 
one seeks states with isospin zero, then the
Finkelstein--Rubinstein signs are all +, and the rotational states
of a tetrahedral Skyrmion have the same spin/parities $J^P$ as above, 
and energies proportional to 
$J(J+1)$. Oxygen-16 can be modelled this way, starting with a Skyrmion
that consists of a tetrahedral cluster of $B=4$ subunits. Recent work 
has gone beyond rigid body quantization and this gives further 
states \cite{HKM}, and a better fit to the experimental spectrum of 
Oxygen-16. Even if Calcium-40 is intrinsically tetrahedral, its 
spectrum will combine vibrational and rotational states.  

Within the lightly bound Skyrme model, the Finkelstein--Rubinstein
signs can be calculated for rigidly rotating clusters of arbitrary 
shape and any baryon number \cite{GHK}, and the spins of some 
low-lying quantum states have been 
determined. There is no doubt that, using rigid body quantization, 
the same spin/parities would be obtained for Calcium-40 as for Oxygen-16 if
one modelled the $B=40$ Skyrmion as a truncated
tetrahedron. It is less clear what would result if one quantized
Skyrmions with higher baryon numbers. Coulomb effects need to be
considered, and more importantly, the related asymmetry between
neutron and proton numbers.  

\section{Wigner's Model and Tetrahedral Symmetry}

We will not review Wigner's model of nuclei \cite{Wig} in detail. Wigner
made the simplyfying assumption of an SU(4) symmetry in nuclear
physics, and treated the four states of the proton or neutron with spin
up or spin down as a fundamental quartet of SU(4). Larger nuclei are
then classified by irreducible representations (irreps) of SU(4). The weight 
diagrams of suitable irreps resemble the truncated 
tetrahedral clusters we have been discussing, and the weight labels
include spin and isospin labels. 

SU(4) is a Lie group of rank 3, so its root lattice and weight lattice are 
three-dimensional \cite{Hum}. The root lattice is an FCC lattice. The weight 
lattice is reciprocal to this, so it is a BCC lattice, and it has four times
as many points. There are four cosets of the root lattice in the
weight lattice, and the weights of each irrep lie in just one of
these. The cosets are shifted FCC lattices, but in thinking about them 
we do not shift the origin. The truncated tetrahedra of interest to us
are all in the coset that contains the weights of the
fundamental 4-dimensional irrep. Further clusters with their centres at the
origin are in other cosets. For example, the $B=13$ cluster mentioned 
earlier is in the root lattice, and the $B=6$ octahedron and the cubic
$B=14$ cluster are in the coset of the 6-dimensional irrep (the vector 
of SO(6)). Their additional symmetry arises from the $\mathbb{Z}_2$ reflection
symmetry of the SU(4) Dynkin diagram.

An important feature of weight diagrams of SU(4) is that typically,
the interior weights in a diagram have multiplicities greater than one.   
The 4-dimensional irrep, with its tetrahedral weight diagram, 
accommodates four nucleons -- one of each type --
and filling the four states gives an alpha particle. This 
is analogous to the $B=4$ tetrahedron in the lightly bound 
Skyrme model modelling an alpha particle. The next irrep 
whose weight diagram has a truncated tetrahedral shape is 
20-dimensional. The shape is the same as the truncated tetrahedron 
modelling Oxygen-16 in the lightly bound Skyrme model, but in 
Wigner's SU(4) model it accommodates 20 nucleons, because
the inner four weights have multiplicity two.

Despite its simplicity, Wigner's model has lasting interest,
and the more sophisticated variants that Wigner discussed in his
original paper, with partial breaking of the SU(4) symmetry, capture
phenomena of physical significance. However, Wigner does not seem to
have argued that his weight diagrams have a spatial interpretation,
despite being three-dimensional. The weight labels are internal 
quantum numbers.

Cook, Dallacasa and collaborators \cite{Coo, CD}, as well as others 
\cite{Eve,Lez}, have rediscovered Wigner's model, and have 
reinterpreted the FCC lattice as a model of the spatial structure of nuclei.  
Nuclei are clusters with at most one nucleon at each lattice site,
but the nucleons acquire labels similar to those of Wigner. The
labels combine the principal quantum number of the isotropic harmonic 
oscillator with the total angular momentum, together with spin and
isospin labels $\pm \frac{1}{2}$. The most stable nuclei, in which 
complete shells of the isotropic harmonic oscillator are filled, have the 
shapes of truncated tetrahedra. Despite these models of nuclei appearing to
be static, individual nucleons have angular momenta that increase as
one moves away from a chosen axis -- which is physically reasonable --
and interestingly, spin up and spin down nucleons occur in complete,
alternating planar layers. Similarly, protons and neutrons (isospin up and 
isospin down) occur in complete, alternating planar layers in an orthogonal
direction. The layer structure is inherited from Wigner's
classification, where it occurs in the three-dimensional weight
space, but Cook et al. argue that it occurs in physical space.

One might criticise the spatial interpretation as having little physical
justification, but it is interesting to compare it with the lightly
bound Skyrme model. The Skyrme model has a physical basis as an
effective field theory of pions, with solitons representing the nucleons. The
truncated tetrahedra arise as particularly strongly
bound arrangements of $B=1$ Skyrmions. A key difference from both
Wigner's model and Cook's reinterpretation is that the $B=1$ Skyrmions 
occur in four distinct orientations, rather than as four
distinct nucleon states. Static Skyrmions are not yet nuclei. Only after
quantization of the complete Skyrmion structure, using rigid body 
quantization or something more sophisticated, does one get a nucleus 
with an overall spin and isospin. 

From the perspective of Skyrmions, it might therefore appear that there is no
spin and isospin layering. However, that is not the case. To see this
one should consider, not a nucleus with spin and isospin zero, like
Calcium-40, but a nucleus with a small net spin, or a small net
isospin. It is a useful approximation to model such nuclei as classically 
spinning or isospinning Skyrmions. Such an approximation can even be
used for $B=1$ Skyrmions. By finding the semi-classical 
approximations to the Adkins, Nappi and Witten quantum states of 
$B=1$ Skyrmions \cite{ANW}, Gisiger and Paranjape \cite{GP} noted 
that protons always spin 
clockwise relative to a particular body axis of the $B=1$ Skyrmion (the 
body axis defined by the neutral pion field), and neutrons always 
spin anticlockwise. The axis is free to point in any direction in
space, which therefore allows protons or neutrons to be spin up or 
down relative to any spatial axis. This classical approximation has
been found useful for studying the collisions of two nucleons in the
Skyrme model \cite{GP,FM}. 

Now suppose a truncated tetrahedral Skyrmion has a small net isospin. 
In one set of planar layers, the $B=1$ Skyrmion constituents occur in two
orientations, but for both of them the body axes point up. In the
alternating set of layers, the constituents again have two
orientations, and for both of them the body axes point down. Because 
of the (classical) isospin, the $B=1$ Skyrmions are all spinning 
clockwise around these body axes, which produces an excess of protons
over neutrons (or anticlockwise, producing an excess of neutrons). 
That means that in the first set of layers, the spins are
all down, and in the second set of layers, the spins are all up (or vice 
versa). Similarly, if there is a net spin aligned with the preferred body 
axes, then the isospins alternate between the layers. This spin and
isospin layering has much similarity to what Cook et al. 
describe, but for the Skyrmions it requires some dynamics, and is
present only in the sense of a quantum superposition if there is no
net spin or isospin, as for example in Calcium-40. 

\section{Remarks on Skyrmion Quantization}

For the Skyrmions that are truncated tetrahedra, with magic baryon 
numbers, low-energy quantum states are probably best
found using collective quantization of the rotational and
isorotational degrees of freedom. Additional states arise from 
collective vibrational modes. The evidence for this comes from 
previous work on the $B=4$ Skyrmion \cite{BBT}, and also the $B=32$
Skyrmion \cite{Fei}, and on the $B=12$ and $B=16$ Skyrmions where
detailed spectra match those found experimentally in Carbon-12 and
Oxygen-16 \cite{LM,HKM}. 

Theoretically, in the shell model, one may interpret a magic nucleus
as having a rather rigid quantum state, because all the available
one-particle states up to some level are occupied, and this rigidity is
enhanced by the short-range nucleon-nucleon repulsion. The Pauli principle
allows one-particle excitations only if a particle is excited to the
next shell up, and this takes considerable energy. Similarly, the
quantum ground state of the Skyrmion, with spin and isospin zero,
involves little relative motion of the $B=1$ Skyrmion constituents. The
$B=1$ Skyrmion locations and orientations are highly organised, as 
in a crystal. Also, similarly as in the shell model, the energy
required to move a single Skyrmion from a complete truncated
tetrahedral cluster up to the next layer is rather large, as
previously mentioned.

But now consider the quantum state of a Skyrmion with baryon number
just one or two greater than a magic number. The Skyrmion will have a
truncated tetrahedral core, to which will be attached one or two 
additional $B=1$ Skyrmions. There is considerable freedom as to where
these additional Skyrmions are. Typically there are
numerous locations in the FCC lattice where one additional Skyrmion can
be attached with 3 bonds, and the energy is almost the same for
all of these. This suggests that the additional Skyrmion should be
treated like a valence nucleon, free to move
in the outer shell it occupies. Rigid body quantization of a particular 
configuration, as considered in \cite{GHK}, is not justified here. 
Similarly, if there are two additional $B=1$ Skyrmions, they are
free to move fairly independently, although 
there is some preference for them to be close together, as this can 
create one additional bond.

The physics of such Skyrmions is therefore quite similar to the 
usual shell model physics of one or two additional nucleons 
interacting with a magic nucleus as core. The additional nucleons are 
fairly free, but there is an important, attractive residual
interaction between them \cite{Cas}. The residual interaction becomes 
more significant if there are three valence nucleons, as it produces a
significant spatial correlation between them, and the simplest 
shell model picture starts to break down. This matches what we 
have seen for lightly bound Skyrmions, where we saw that it was favourable 
to attach three $B=1$ Skyrmions in the form of a triangle, because this adds 
12 extra bonds. The three Skyrmions are strongly correlated spatially. 

There remain some challenges for the Skyrme model here. It is
important to see if the quantization of a single $B=1$ Skyrmion outside
a core leads to a strong spin-orbit coupling. The orientation of the
Skyrmion varies with its location, so it is plausible that as it moves
across the surface of the core it has to spin too. An analysis of this
coupling has been carried out in the simpler Baby Skyrme model in two 
dimensions \cite{HM}, but not yet in the context of the three-dimensional
model. One should probably allow the $B=1$ Skyrmion to move freely around the
core, but a possible simplification is to constrain the $B=1$ Skyrmion
to occupy one of the FCC lattice sites (of which there are just a
finite number in the layer outside a truncated tetrahedral core). The 
Hamiltonian for the $B=1$ Skyrmion would then be a hopping
Hamiltonian, as used frequently in condensed matter contexts. A 
further challenge is to allow for rotations of the core. It is 
presumably necessary to parametrise the orientation of the core using 
continuous coordinates (Euler angles) even if the $B=1$ Skyrmion 
outside is treated as hopping.    

\section*{Appendix: Rectangular Bipyramids and Octahedra}

In this paper, we mainly considered the Skyrmions obtained from a
tetrahedron with $2N$ baryons along an edge, truncated by removing
four tetrahedra with $N-1$ baryons along an edge.
The eight faces of such a Skyrmion alternate between 
equilateral triangles of baryons and slightly larger hexagons with one 
bond along each short edge. To pass from one truncated tetrahedron to the
next, it is sufficient to attach the next larger equilateral triangle 
of baryons to each of the four hexagonal faces. Each pair of these 
triangles is joined by a single bond.

If just two of these equilateral triangles are attached, then the
baryon number is half-way between that of the truncated 
tetrahedron one starts with, and the next one. The number of bonds is 
just two less than half-way between, because the two triangles are 
joined by a single bond, whereas if all four triangles are attached, they 
are joined by 6 bonds. The baryon number sequence obtained this way 
is therefore $B = 10 \,, 28 \,, 60 \,, 110 \,, 182 \,, 280$ and the 
bond numbers (binding energies) are $E = 25 \,, 97 \,, 241 \,, 481 
\,, 841 \,, 1345$. The binding energies per baryon are 
$E/B = 2.5 \,, 3.46 \,, 4.02 \,, 4.37 \,, 4.62 \,, 4.80$. We do not 
give the algebraic formulae, but these are easily deduced from those 
for the truncated tetrahedra. They are not especially simple.

The shapes of these clusters are rather elegant. They are rectangular 
bipyramids, with ${\rm D}_{2h}$ symmetry. This is best seen through 
their slicings into rectangles of baryons. For example, the $B=28$ 
bipyramid is sliced into $2 + 6 + 12 + 6 + 2$ baryons. 
The rectangles have sides that differ by one baryon, and they
all have the same orientation (see Figure 3). For comparison, 
the $B=16$ and $B=40$ truncated tetrahedra slice into $2 + 6 + 6 + 2$ 
baryons and $2 + 6 + 12 + 12 + 6 + 2$ baryons, respectively (see
Figures 1 and 2). Here, the lower rectangles (the second half of 
the sequence) are rotated by $\pi/2$ relative to the upper rectangles.

\begin{figure}[!ht]
\centering
\includegraphics[width=10.5cm]{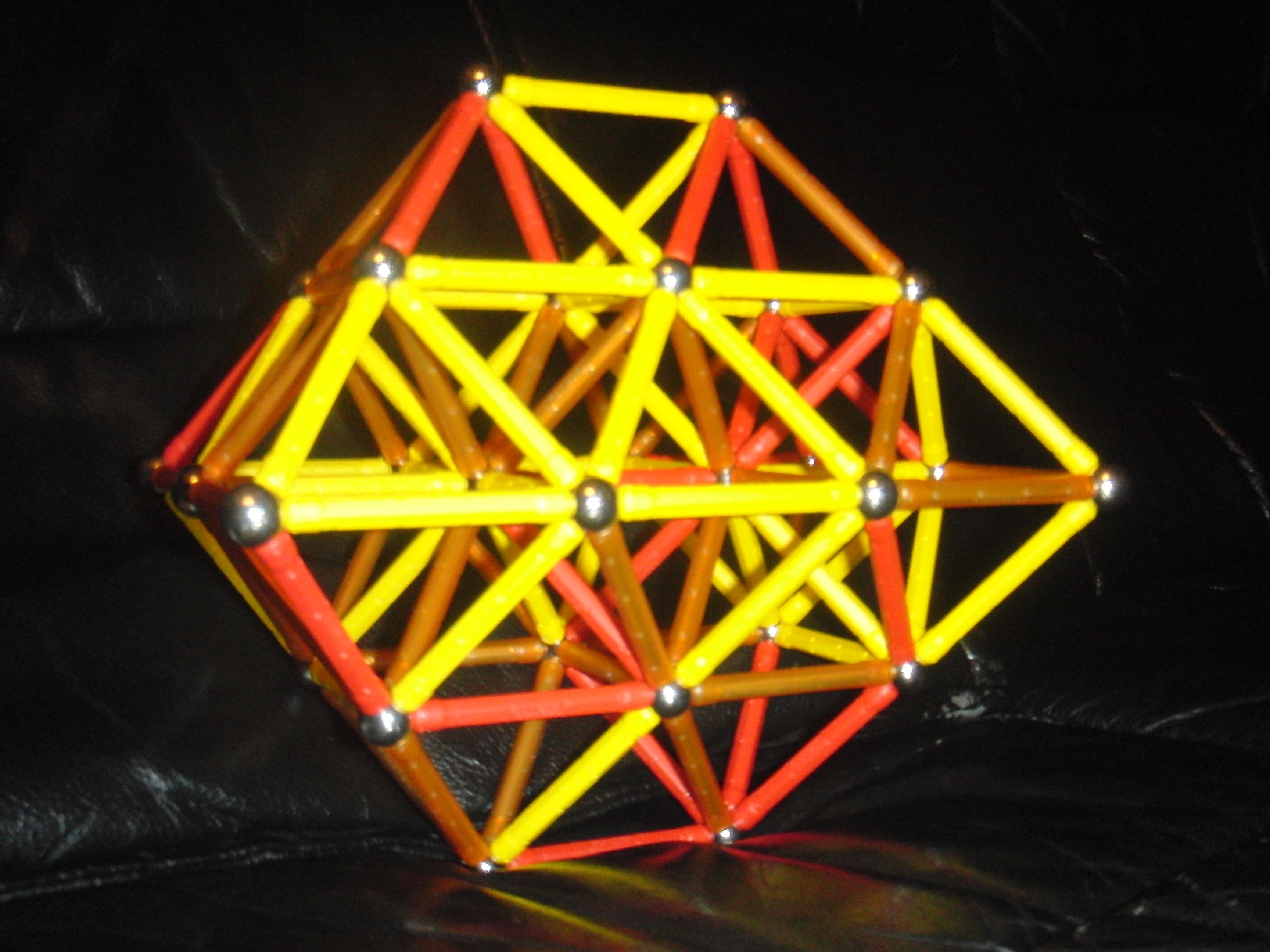}
\caption{$B=28$ rectangular bipyramid.}
\label{fig3}
\end{figure}

These bipyramids are interesting because of their high bond numbers.
The 25 bonds of the $B=10$ bipyramid is the maximum possible
for this baryon number \cite{GHK}. The $B=28$ example has one more
bond than the tetrahedrally symmetric cluster obtained by attaching 
four $B=3$ triangles to the $B=16$ truncated tetrahedron, as discussed
in Section 2.

We now turn to more symmetric Skyrmions, with cubic symmetry. There
is an infinite sequence of complete, pure octahedra. They alternate
between having a baryon at the centre and not. The sequences of baryon
numbers, bond numbers (binding energies), and the energies per baryon are
$B = 1 \,, 6 \,, 19 \,, 44 \,, 85 \,, 146 \,, 231$, 
$E = 0 \,, 12 \,, 60 \,, 168 \,, 360 \,, 660 \,, 1092$, and
$E/B = 0 \,, 2 \,, 3.16 \,, 3.82 \,, 4.24 \,, 4.52 \,, 4.73$.
The baryon numbers are found by slicing the octahedra into squares.
For example, the $B=85$ octahedron has square slices $1 + 4 +
9 + 16 + 25 + 16 + 9 + 4 + 1$. The bond numbers are high. For example, 
the $B=85$ octahedron has 24 bonds more than the $B=80$ truncated 
tetrahedron. Generally, the numbers above slightly exceed those in the 
sequences for the truncated tetrahedra (starting with $B=0$,
$E=0$). At each step, the difference in baryon number increases by 1, 
and the difference in bond number increases by 6.  
The algebraic formulae for the baryon number and bond number of an 
octahedron with $N$ baryons along an edge are
\be
B = \frac{1}{3}(2N^2 + 1)N \,, \qquad
E = 2(2N - 1)N(N - 1) \,.
\ee
Note that $B = T_{2N-1} - 4T_{N-1}$, because an
octahedron can be obtained by suitably truncating a complete
tetrahedron with $2N-1$ baryons along an edge. For example, the $B=19$
octahedron is a truncation of the $B=35$ tetrahedron. (A minimal
truncation of the $B=35$ tetrahedron gives a $B=31$ truncated
tetrahedron with $E/B = 3.48$, a high value.)  

For the larger octahedra, an even higher binding energy per baryon is
achieved by truncating the six corners, removing one baryon from each. 
The truncation removes six baryons and 24 bonds, leaving six square 
faces and eight hexagonal faces. The 
sequences of baryon numbers and bond numbers (binding energies) for
the truncated octahedra are 
$B = 13 \,, 38 \,, 79 \,, 140 \,, 225$ and
$E = 36 \,, 144 \,, 336 \,, 636 \,, 1068$,
and the binding energies per baryon are
$E/B = 2.77 \,, 3.79 \,, 4.25 \,, 4.54 \,, 4.75$.
The first of these corresponds to a truncated $B=19$ octahedron, with a
central baryon and its twelve nearest neighbours. Note that the
truncated octahedron with $B=140$ has 6 bonds more than the
truncated tetrahedron with the same baryon number.

The truncated octahedra, starting with $B=38$ (see Figure 4), 
have the following interesting property. The vertices have coordination 
number 6, so removing a vertex baryon to infinity requires breaking 
6 bonds. This is the maximum possible for a convex
vertex, which is mathematically related to the fact that the
vectors from a vertex baryon to its six nearest neighbours possess the
same geometry as the six positive roots of SU(4). Note also that the 
surface of a truncated octahedron is topologically a sphere, but 
its curvature is concentrated at the vertices. The curvature is
particularly small for these special vertices, and to compensate, 
there are 24 of them.

\begin{figure}[!ht]
\centering
\includegraphics[width=10.5cm]{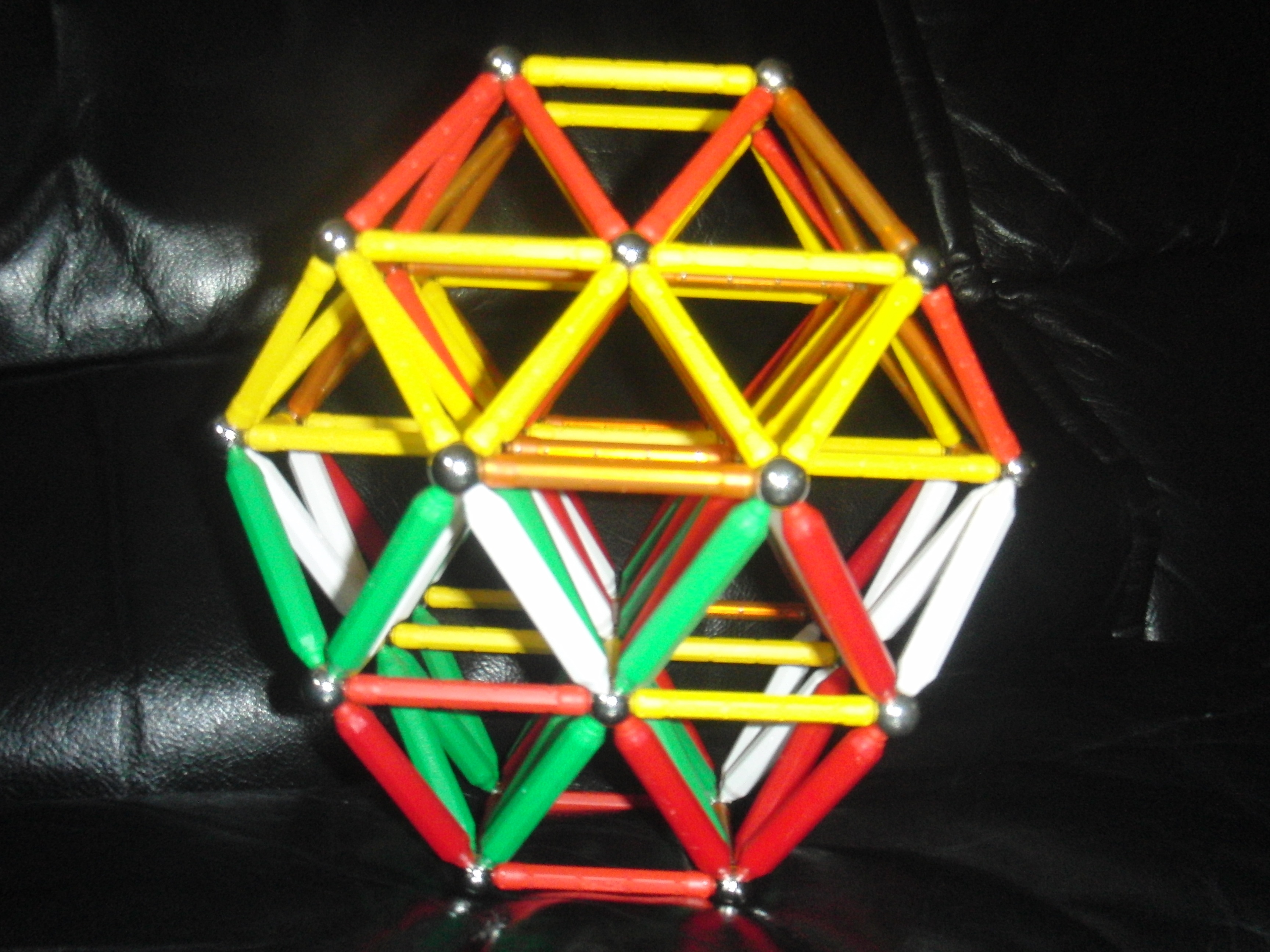}
\caption{$B=38$ truncated octahedron.}
\label{fig4}
\end{figure}

It is tempting to think of the complete octahedra and truncated 
octahedra as magic, but the calculations in \cite{GHK} show that, at
least for the examples with baryon numbers 6, 13 and
19, the binding energies are not exceptionally large when the
interactions between all of the $B=1$ Skyrmions are allowed for. This
appears to be because the $B=1$ Skyrmions are not distributed between 
the four orientations as equally as possible.

A more radical truncation of an octahedron, removing five baryons from
each corner, is not optimal until the baryon numbers becomes larger than   
those relevant to nuclei.

\section*{Acknowledgements}

I am grateful to Martin Speight for discussions, and for clarifying 
that one of the truncated tetrahedra has 150 bonds. I also thank Derek
Harland for helpful comments. The figures are photos of lightly bound 
Skyrmions, modelled using the magnetic building sets SUPERMAG-Maxi and 
GEOMAG. This research is partly supported by STFC grant ST/L000385/1.

\end{document}